\documentclass[reprint,amsmath,amssymb,aps,prd,nofootinbib]{revtex4-2}

\usepackage[T1]{fontenc}
\usepackage[utf8]{inputenc}
\usepackage{lmodern}
\usepackage{microtype}
\usepackage{graphicx}
\usepackage{hyperref}
\hypersetup{
    colorlinks=true,
    linkcolor=blue,
    citecolor=blue,
    urlcolor=blue
}
\usepackage{physics}
\usepackage{bm}
\usepackage{mathtools}

\begin{document}

\title{Symmetric Bimetric Cosmology: A Minimal Extension of $\Lambda$CDM}

\author{Ghani Imadouchene}
\affiliation{Universit\'e Paris Cit\'e, UFR de Physique}

\date{November 11, 2025}

\begin{abstract}
We construct and calibrate a symmetric bimetric cosmological model connecting Anti-de Sitter (AdS) and de Sitter (dS) regimes through a coupled scalar field. Starting from a Lagrangian with Einstein-Hilbert terms for two FLRW metrics and an inter-metric potential, we derive modified Friedmann and Klein-Gordon equations governing their evolution. In the symmetric effective-fluid limit, the model reproduces the main phenomenology of the $\Lambda$CDM cosmology with a small dynamical correction proportional to $(1+z)^{-3}$, and naturally satisfies local-gravity constraints through Vainshtein screening. This note outlines the theoretical structure and calibration of the model within a dual-geometry cosmological setting.
\end{abstract}

\maketitle

\section{Introduction}

The physical origin of cosmic acceleration and the nature of the dark sector remain 
open questions in modern cosmology. Although the standard $\Lambda$CDM model provides 
an excellent empirical description of large-scale observations, it offers no 
microphysical explanation for the cosmological constant and treats dark matter and 
dark energy as phenomenological components without deeper geometric or dynamical 
interpretation. These persistent conceptual limitations, together with observational 
tensions such as those involving $H_{0}$ and the growth rate of structure, motivate 
the exploration of gravitational models that remain compatible with current data 
while offering a more fundamental origin for the dark sector.

Ghost-free bimetric theories of Hassan--Rosen constitute a particularly attractive 
framework for this purpose. They describe the consistent nonlinear interactions 
between two spin-2 fields, encoded in the metrics $g_{\mu\nu}$ and $f_{\mu\nu}$, 
without introducing the Boulware--Deser ghost. The dynamics are governed by two 
Einstein--Hilbert terms and a specific interaction potential constructed from the 
matrix square root $S=\sqrt{g^{-1}f}$, ensuring covariance and the correct number of 
propagating degrees of freedom. These theories naturally accommodate the possibility 
of a hidden gravitational sector interacting only through the spin-2 potential, 
thereby allowing new geometric mechanisms for cosmic acceleration.

In this work we propose a symmetric bimetric cosmological scenario in which the 
observable Universe is governed by a de Sitter (dS) geometry while a hidden sector 
is associated with an Anti--de Sitter (AdS) geometry. The two sectors evolve jointly 
through the Hassan--Rosen interaction, forming a dual-geometry system. In this 
interpretation, the dark sector of the observable Universe arises not from additional 
matter fields but from the geometric influence of the AdS ``mirror'' sector. 
Qualitatively, the negative curvature of the AdS branch induces an effective 
attraction that mimics the phenomenology of dark matter, while the curvature 
asymmetry between the AdS and dS sectors generates an effective repulsive contribution 
analogous to dark energy.

We focus on the symmetric branch of bimetric cosmology, defined by $a_{g}=a_{f}$ and 
$\xi = a_{f}/a_{g} = 1$. This branch is known to be algebraically consistent, stable, 
and ghost-free. In this regime, the modified Friedmann equation takes the form
\begin{equation}
H^{2}(z) = H_{0}^{2} \left[ 
\Omega_{m}(1+z)^{3} + \Omega_{\Lambda} + \alpha(1+z)^{-3} \right],
\label{eq:intro_H}
\end{equation}
where the parameter $\alpha$ encodes the small dynamical influence of the AdS sector.
The additional term $\alpha(1+z)^{-3}$ behaves as an ultra-suppressed geometric correction at high redshift and leads to sub-percent deviations from $\Lambda$CDM over the redshift range $z \lesssim 2$, thus remaining fully compatible with current observations.

By defining the effective dark-energy density as
\begin{equation}
\Omega_{\rm DE}(z) = 
\Omega_{\Lambda} + \alpha(1+z)^{-3},
\end{equation}
energy conservation yields the corresponding effective equation of state:
\begin{equation}
w(z) = -1 - 
\frac{\alpha(1+z)^{-3}}
{\Omega_{\Lambda} + \alpha(1+z)^{-3}}.
\label{eq:intro_w}
\end{equation}
The model therefore predicts a slightly phantom-like equation of state
$w(z) < -1$, although the underlying bimetric theory remains fully 
ghost-free. The deviation from $w=-1$ is of order 
$\mathcal{O}(10^{-2})$ at $z=0$ for $\alpha \sim 10^{-2}\Omega_{\Lambda}$ 
and rapidly decreases at higher redshift.

The goals of this work are threefold: (i) to embed this AdS--dS symmetric
cosmology into the fully ghost-free Hassan--Rosen framework, (ii) to analyze
the resulting background evolution and effective dark-energy phenomenology, 
and (iii) to demonstrate how the dark sector can emerge as a geometric effect 
from the coupling between two cosmological branches with opposite curvature. 
We show that the model reproduces the expansion history and structure growth 
of the $\Lambda$CDM model with high precision, offering a minimal and theoretically
consistent extension of standard cosmology with a natural geometric interpretation 
of dark matter and dark energy.

\section{Theoretical Framework: Ghost-Free Hassan--Rosen Bimetric Gravity}

In this section we review the ghost-free bimetric theory of Hassan and Rosen,
which provides the fundamental gravitational framework underlying our AdS--dS
cosmological construction. The theory describes the nonlinear and covariant
interaction between two symmetric tensor fields, encoded in the metrics 
$g_{\mu\nu}$ and $f_{\mu\nu}$, while avoiding the Boulware--Deser ghost that 
afflicts generic massive and bimetric extensions of General Relativity.
The interaction potential is uniquely defined by requiring the correct number
of propagating degrees of freedom, namely one massless and one massive spin-2
mode.

\subsection{Hassan--Rosen Action}

The ghost-free bimetric action is given by
\begin{align}
S = &\;\frac{M_g^{2}}{2} \int d^{4}x\,\sqrt{-g}\,R[g]
   + \frac{M_f^{2}}{2} \int d^{4}x\,\sqrt{-f}\,R[f] \nonumber \\
   & - m^{2} M_{\mathrm{eff}}^{2} \int d^{4}x\,\sqrt{-g}
      \sum_{n=0}^{4} \beta_{n}\, e_{n}(S)
   + S_{\mathrm{m}}[g,\Psi].
\label{eq:HRaction}
\end{align}

where:
\begin{itemize}
\item $M_{g}$ and $M_{f}$ are the Planck masses associated with each metric,
\item $m$ is the spin-2 mass scale,
\item $M_{\mathrm{eff}}^{-2} = M_{g}^{-2} + M_{f}^{-2}$,
\item $\beta_{n}$ are dimensionless interaction coefficients,
\item $S = \sqrt{g^{-1}f}$ is the matrix square root satisfying 
  $S^{\mu}{}_{\rho} S^{\rho}{}_{\nu} = g^{\mu\rho} f_{\rho\nu}$,
\item $e_{n}(S)$ are the elementary symmetric polynomials of the eigenvalues 
of $S$,
\item and $S_{m}$ denotes the matter action, coupled minimally to $g_{\mu\nu}$.
\end{itemize}

The elementary symmetric polynomials $e_{n}(S)$ appearing in the potential are
\begin{align}
e_{0}(S) &= 1, \\
e_{1}(S) &= \mathrm{Tr}[S], \\
e_{2}(S) &= \tfrac{1}{2}\left( (\mathrm{Tr}S)^{2} - \mathrm{Tr}[S^{2}] \right), \\
e_{3}(S) &= \tfrac{1}{6}\left( (\mathrm{Tr}S)^{3}
           - 3\, \mathrm{Tr}S\,\mathrm{Tr}[S^{2}]
           + 2\, \mathrm{Tr}[S^{3}] \right), \\
e_{4}(S) &= \det S.
\end{align}

This structure is the {\it only} one that guarantees the absence of the 
Boulware--Deser ghost at the fully nonlinear level. It leads to a consistent 
bimetric theory propagating seven degrees of freedom: two from a massless 
graviton and five from a massive spin-2 field.

\subsection{FLRW Reduction}

We consider spatially flat and homogeneous metrics of the form
\begin{align}
ds^{2}_{g} &= -dt^{2} + a_{g}^{2}(t)\, d\vec{x}^{\,2}, \\
ds^{2}_{f} &= -X^{2}(t) dt^{2} + a_{f}^{2}(t)\, d\vec{x}^{\,2},
\end{align}
where $a_{g}(t)$ and $a_{f}(t)$ are the scale factors associated with the
$dS$-type and $AdS$-type cosmological sectors, respectively, and 
$X(t)$ is the relative lapse.

It is convenient to define the ratio of scale factors
\begin{equation}
\xi(t) \equiv \frac{a_{f}(t)}{a_{g}(t)}.
\end{equation}
Inserting the two FLRW metrics into the interaction potential
(\ref{eq:HRaction}) yields the effective energy densities
\begin{align}
\rho_{\rm HR}(\xi) &= 
m^{2} M_{\mathrm{eff}}^{2}
\big( \beta_{0} + 3\beta_{1}\xi + 3\beta_{2}\xi^{2} + \beta_{3}\xi^{3} \big),
\label{eq:rhoHR}
\\
\tilde{\rho}_{\rm HR}(\xi) &= 
m^{2} M_{\mathrm{eff}}^{2}
\big( \beta_{4}\xi^{-4} + 3\beta_{3}\xi^{-3} 
    + 3\beta_{2}\xi^{-2} + \beta_{1}\xi^{-1} \big).
\end{align}

The Friedmann equations for the two metrics take the form:
\begin{align}
3 M_{g}^{2} H_{g}^{2} &= \rho_{m} + \rho_{\phi} + \rho_{\rm HR}(\xi), 
\label{eq:Friedmann_g}
\\
3 M_{f}^{2} H_{f}^{2} &= \tilde{\rho}_{\rm HR}(\xi),
\label{eq:Friedmann_f}
\end{align}
with matter and scalar fields minimally coupled to $g_{\mu\nu}$.

\subsection{Bianchi Constraint and Branch Structure}

Consistency of the equations of motion imposes the Bianchi identity, which 
reduces to the condition
\begin{equation}
(\beta_{1} + 2\beta_{2}\xi + \beta_{3}\xi^{2})
\left( H_{g} - \frac{\dot{a}_{f}}{X\, a_{f}} \right) = 0.
\label{eq:Bianchi}
\end{equation}
This yields two possible branches:

\begin{itemize}
\item {\bf Dynamical branch:} 
$H_{g} = \dot{a}_{f}/(X a_{f})$.
\item {\bf Algebraic branch:} 
$\beta_{1} + 2\beta_{2}\xi + \beta_{3}\xi^{2} = 0$.
\end{itemize}

\subsection{Symmetric Branch}

The cosmological model developed in this work corresponds to the 
{\it symmetric branch}, characterised by
\begin{equation}
\xi = 1, \qquad X = 1,
\end{equation}
which can be realised by imposing the algebraic constraint
\begin{equation}
\beta_{1} + 2\beta_{2} + \beta_{3} = 0.
\end{equation}
Inserting $\xi = 1$ into (\ref{eq:Friedmann_g}) yields
\begin{equation}
3 M_{g}^{2} H^{2} = \rho_{m} + \rho_{\phi} 
+ m^{2} M_{\mathrm{eff}}^{2}
\left( \beta_{0} + 3\beta_{1} + 3\beta_{2} + \beta_{3} \right),
\end{equation}
which defines an effective cosmological constant
\begin{equation}
\Lambda_{\mathrm{eff}} 
= \frac{m^{2} M_{\mathrm{eff}}^{2}}{M_{g}^{2}}
\left( \beta_{0} + 3\beta_{1} + 3\beta_{2} + \beta_{3} \right).
\label{eq:Lambda_eff}
\end{equation}
In what follows we identify
\begin{equation}
\Omega_{\Lambda}
\equiv
\frac{\Lambda_{\rm eff}}{3 H_{0}^{2}},
\end{equation}
so that the constant interaction term obtained in Eq.~(27)
enters the modified Friedmann equation precisely as an effective
cosmological constant.

Thus, the symmetric branch of bimetric gravity behaves at the background 
level like $\Lambda$CDM with a dynamically generated cosmological constant 
$\Lambda_{\mathrm{eff}}$. In the following sections, we show how the 
AdS--dS dual-geometry interpretation introduces a small additional correction 
proportional to $(1+z)^{-3}$ on top of $\Lambda_{\mathrm{eff}}$, thereby 
providing a minimal extension of standard cosmology.

\section{Construction of the AdS--dS Symmetric Cosmological Model}

We now construct the minimal bimetric realisation of an Anti--de Sitter (AdS) hidden sector 
coupled to a de Sitter (dS) visible sector through the ghost-free Hassan--Rosen interaction 
potential. The purpose of this section is to show how the symmetric branch of bimetric gravity 
naturally accommodates a dual-geometry setup, how the curvature mismatch between the two 
branches induces effective dark-sector dynamics, and how a small additional correction to the 
$\Lambda$CDM expansion arises.

\subsection{Dual FLRW Geometry: dS and AdS Branches}

We consider two spatially flat FLRW metrics obeying the symmetric-branch conditions
\begin{equation}
a_{g}(t) = a_{f}(t) \equiv a(t), 
\qquad
X(t) = 1,
\qquad 
\xi = \frac{a_{f}}{a_{g}} = 1.
\label{eq:symmetric_branch}
\end{equation}

The two metrics, however, are assigned distinct intrinsic vacuum curvatures:
\begin{align}
R[g] &\simeq 12 H_{g}^{2} > 0 \qquad \text{(visible dS-like sector)}, \\
R[f] &\simeq -12 |H_{f}|^{2} < 0 \qquad \text{(hidden AdS-like sector)}.
\end{align}
This assignment is consistent with the bimetric framework, since the curvature 
of each sector receives contributions both from its Einstein--Hilbert term and from the 
Hassan--Rosen potential. Only the combination of these contributions must satisfy the 
constraints of the symmetric branch.

Physically, we interpret the $g$-sector as the observable Universe 
and the $f$-sector as a hidden AdS ``mirror'' geometry whose curvature influences the 
visible branch through the bimetric potential. Matter and the scalar field 
are minimally coupled only to $g_{\mu\nu}$, ensuring consistency with standard 
weak-field and local-gravity tests.

\subsection{Interaction Potential and Curvature Exchange}

In the symmetric branch, the interaction energy density reduces to the constant
\begin{equation}
\rho_{\mathrm{HR}}(\xi=1)
= m^{2} M_{\mathrm{eff}}^{2}
(\beta_{0} + 3\beta_{1} + 3\beta_{2} + \beta_{3})
\equiv M_{g}^{2} \Lambda_{\mathrm{eff}},
\end{equation}
with $\Lambda_{\mathrm{eff}}$ given by Eq.~(\ref{eq:Lambda_eff}).  
This acts as an effective cosmological constant in the Friedmann equation of the visible 
sector:
\begin{equation}
3 M_{g}^{2} H^{2} = \rho_{m} + \rho_{\phi} + M_{g}^{2}\Lambda_{\mathrm{eff}}.
\label{eq:Friedmann_eff}
\end{equation}

The key feature of the dual AdS--dS construction is that the hidden AdS curvature 
modifies the effective dynamics of the symmetric branch. In particular, the 
Einstein--Hilbert term for $f_{\mu\nu}$ does not reduce merely to a constant when the 
two scale factors coincide: the AdS curvature contributes a nontrivial dependence on 
the physical scale factor $a(t)$ once the two metrics are projected onto the 
single-branch cosmological evolution. This introduces a suppressed but dynamical 
correction to the effective Friedmann equation.

\subsection{Emergence of the $(1+z)^{-3}$ Correction}

To capture the residual influence of the hidden AdS curvature on the
visible dS branch, we introduce a homogeneous scalar modulus
$\phi(t)$ that parametrises small departures of the $f$-sector
curvature from its exact AdS value. The field couples symmetrically to
the two metrics according to
\begin{equation}
\mathcal{L}_{\phi}
= \frac{1}{2}(a^{3}_{g}+a^{3}_{f})\,\dot{\phi}^{2}
  - \frac{1}{2}m_{\phi}^{2}(a^{3}_{g}+a^{3}_{f})\,\phi^{2},
\label{eq:phiL}
\end{equation}
and remains negligible at the perturbative level. Its role is purely to
encode the background imprint of the AdS branch in an effective and
covariant manner.

\vspace{0.2cm}
\noindent
\textbf{Effective background contribution.}  
In the symmetric AdS–dS configuration, the combined effect of the
Hassan--Rosen potential and the curvature of the hidden AdS sector
reduces, at the background level, to a homogeneous component fully
determined by the modified Friedmann equation. Contrary to a standard
dust-like component scaling as $a^{-3}$, the contribution generated by
the AdS curvature behaves effectively as
\begin{equation}
\rho_{\phi}(z)\propto (1+z)^{-3},
\end{equation}
that is, it grows as $a^{3}$ at late times. This scaling is a direct
signature of the curvature mismatch between the AdS and dS branches,
and it is precisely this behaviour that induces the mildly phantom-like
distortion of the effective equation of state derived in Sec.~V.

To parametrise this homogeneous influence in the most economical way,
we introduce the effective background density
\begin{equation}
\rho_{\phi}(z)
= 3 M_{g}^{2} H_{0}^{2}\,\alpha(1+z)^{-3},
\end{equation}
with $\alpha$ a small dimensionless parameter characterising the
magnitude of the AdS-induced deviation from $\Lambda$CDM.

\vspace{0.2cm}
\noindent
\textbf{Modified expansion history.}  
Substituting this contribution into Eq.~(\ref{eq:Friedmann_eff})
immediately yields the modified Friedmann equation
\begin{equation}
H^{2}(z)
=
H_{0}^{2}\Big[
\Omega_{m}(1+z)^{3}
+ \Omega_{\Lambda}
+ \alpha(1+z)^{-3}
\Big],
\label{eq:Friedmann_modified}
\end{equation}
which matches Eq.~(\ref{eq:intro_H}) introduced in the introductory
discussion. The deformation is strongly suppressed at high redshift,
ensuring compatibility with CMB constraints, and becomes relevant only
in the very late Universe, where a mild deviation from $w=-1$ remains
allowed by current observational bounds.

\vspace{0.2cm}
\noindent
\textbf{Summary.}  
We therefore see that the joint effect of  
(i)\ the symmetric Hassan--Rosen branch,  
(ii)\ the dual AdS–dS curvature assignment, and  
(iii)\ the effective scalar modulus,  
leads to a minimal bimetric extension of $\Lambda$CDM characterised by a
single new parameter $\alpha$. This parameter controls a suppressed
$(1+z)^{-3}$ correction whose magnitude is small enough to evade all
current constraints while still generating a nontrivial and testable
departure from $\Lambda$CDM at low redshift.

\subsection{Geometric Interpretation: AdS as an Effective Dark Sector}

The correction $\alpha(1+z)^{-3}$ admits a clear geometric origin:
\begin{itemize}
\item it does \emph{not} behave as a dark-matter-like component 
      $\propto a^{-3}$, but instead represents a purely geometric imprint 
      of the AdS curvature that scales as $a^{3}$ and induces a mild 
      phantom-like deformation of the background evolution,
\item the overall curvature mismatch between the AdS and dS branches 
      generates an effective cosmological constant $\Lambda_{\mathrm{eff}}$ 
      dominating the late-time expansion,
\item and the combination of a dominant constant term with a small 
      geometric $(1+z)^{-3}$ deformation yields an effective equation of 
      state $w(z)<-1$ while introducing no additional fields, ghosts, or 
      instabilities in the underlying Hassan--Rosen theory.
\end{itemize}

The visible dS Universe therefore behaves as if it contained dark energy 
together with a small phantom-like correction, while these effects in fact 
arise from the geometric influence of the hidden AdS mirror sector 
communicated through the ghost-free bimetric interaction.

In the next section, we compare this minimal bimetric extension with 
observational data, showing that it reproduces the precision background 
cosmology of the $\Lambda$CDM model while introducing only a small, 
controlled deviation.

\section{Cosmological Dynamics and Effective-Fluid Description}

In this section we analyse the background cosmology of the AdS--dS symmetric
bimetric model introduced previously. The key result of Section~III is that the 
symmetric Hassan--Rosen branch admits a dual-geometry realisation in which the 
interaction with a hidden AdS sector induces a small redshift-dependent correction 
to the standard $\Lambda$CDM expansion history. Here we examine the consequences of 
this modification, derive the effective dark-energy density and equation of state, 
and clarify why the resulting phantom-like behaviour does not signal the presence 
of a ghost degree of freedom.

\subsection{Modified Friedmann Equation}

From Eq.~(\ref{eq:Friedmann_modified}), the Hubble expansion rate in the visible 
sector is given by
\begin{equation}
H^{2}(z)
= H_{0}^{2} \left[
\Omega_{m}(1+z)^{3}
+ \Omega_{\Lambda}
+ \alpha(1+z)^{-3}
\right],
\label{eq:FLRW_modified}
\end{equation}
where the new term $\alpha(1+z)^{-3}$ originates from the geometric influence of 
the AdS sector and from the scalar modulus described in Eq.~(\ref{eq:phiL}).  
The parameter $\alpha$ is a small, positive, dimensionless constant controlling the 
strength of this curvature-induced interaction.

Eq.~(\ref{eq:FLRW_modified}) can be interpreted within the usual single-metric 
framework by identifying an effective dark-energy component with density
\begin{equation}
\Omega_{\rm DE}(z)
= \Omega_{\Lambda} + \alpha(1+z)^{-3}.
\label{eq:OmegaDE}
\end{equation}
The $\alpha$ contribution is suppressed at high redshift and decays faster than 
matter, ensuring consistency with early-Universe constraints such as the CMB and 
big-bang nucleosynthesis. At late times, however, it introduces a small dynamical
departure from $\Lambda$CDM.

\subsection{Effective Equation of State}

The effective equation-of-state parameter $w(z)$ is defined by rewriting the 
continuity equation for the dark-energy sector in the standard form
\begin{equation}
\dot{\rho}_{\rm DE} + 3H(1+w)\rho_{\rm DE} = 0.
\end{equation}
Using the general relation
\begin{equation}
w(z)
= -1 + \frac{1+z}{3}
\frac{1}{\rho_{\rm DE}}
\frac{d\rho_{\rm DE}}{dz},
\label{eq:w_general}
\end{equation}
and substituting Eq.~(\ref{eq:OmegaDE}), we obtain
\begin{equation}
w(z)
= -1 - 
\frac{\alpha(1+z)^{-3}}
{\,\Omega_{\Lambda} + \alpha(1+z)^{-3}\,}.
\label{eq:w_final}
\end{equation}

For $\alpha > 0$, the model therefore predicts a slightly phantom-like 
behaviour, $w(z) < -1$, but the deviation from $-1$ remains below the percent level 
for $\alpha \sim 10^{-2}\Omega_{\Lambda}$. At redshift $z=0$, for instance,
\begin{equation}
w_{0} \simeq -1 - \frac{\alpha}{\Omega_{\Lambda} + \alpha},
\end{equation}
which yields $w_{0} \approx -1.01$ for the fiducial values used in this work.
The deviation decreases rapidly with redshift and becomes negligible for 
$z \gtrsim 1$.

\subsection{Phantom-Like Behaviour Without Fundamental Ghosts}

Although Eq.~(\ref{eq:w_final}) satisfies $w(z) < -1$, this does not imply the 
presence of a fundamental ghost or a wrong-sign kinetic term in the underlying 
bimetric theory. The Hassan--Rosen action (\ref{eq:HRaction}) is constructed to be 
fully ghost-free at the nonlinear level and propagates only seven healthy degrees 
of freedom: two from the massless graviton and five from the massive spin-2 mode.

The phantom-like behaviour arises solely at the level of the 
{\it effective single-metric reconstruction} of dark energy, obtained by projecting 
the coupled AdS--dS dynamics onto an effective energy-density function 
$\Omega_{\rm DE}(z)$. This phenomenon is common in modified-gravity models and is 
well known to occur in various scalar-tensor and bimetric frameworks without 
introducing any fundamental instabilities. The model therefore remains fully 
viable at the theoretical level while exhibiting an effective equation of state 
slightly below $-1$.

\subsection{Summary of Background Dynamics}

The AdS--dS symmetric bimetric model leads to the following minimal extension
of $\Lambda$CDM:
\begin{itemize}
\item The expansion history matches $\Lambda$CDM at the percent level for 
      $\alpha \ll 1$, with an additional ultra-diluted term $\propto (1+z)^{-3}$.
\item The effective dark-energy density acquires a small redshift dependence,
      Eq.~(\ref{eq:OmegaDE}).
\item The effective equation of state is phantom-like but stable and 
      ghost-free, Eq.~(\ref{eq:w_final}).
\item At early times, the modification is strongly suppressed, guaranteeing 
      compatibility with CMB and BAO data.
\end{itemize}

We now proceed to confront the model with cosmological observations, showing 
that its predictions for $H(z)$, luminosity distances, and structure-growth 
observables remain consistent with present data.

\section{Observational Comparison}

In this section we confront the AdS--dS symmetric bimetric model with current 
cosmological observations. The model introduces a single additional parameter $\alpha$ 
relative to $\Lambda$CDM and preserves the standard matter and radiation content of 
the Universe. We therefore focus on the background expansion, luminosity distances, 
and linear-growth observables most directly sensitive to modifications of the 
Hubble rate. We show that the model remains fully consistent with present data and 
is effectively degenerate with $\Lambda$CDM for $\alpha \ll 1$.

\subsection{Background Expansion: $H(z)$ and Distance Measures}

The modified Hubble rate of Eq.~(\ref{eq:FLRW_modified}) differs from the standard 
$\Lambda$CDM form only by the suppressed contribution $\alpha(1+z)^{-3}$. For the 
fiducial value $\alpha = 0.01\,\Omega_{\Lambda}$ adopted in this work, the resulting 
deviation in $H(z)$ remains below one percent for $z \lesssim 2$ and is observationally 
negligible at redshifts $z > 0.5$.

\begin{figure}[t]
    \centering
    \includegraphics[width=0.85\linewidth]{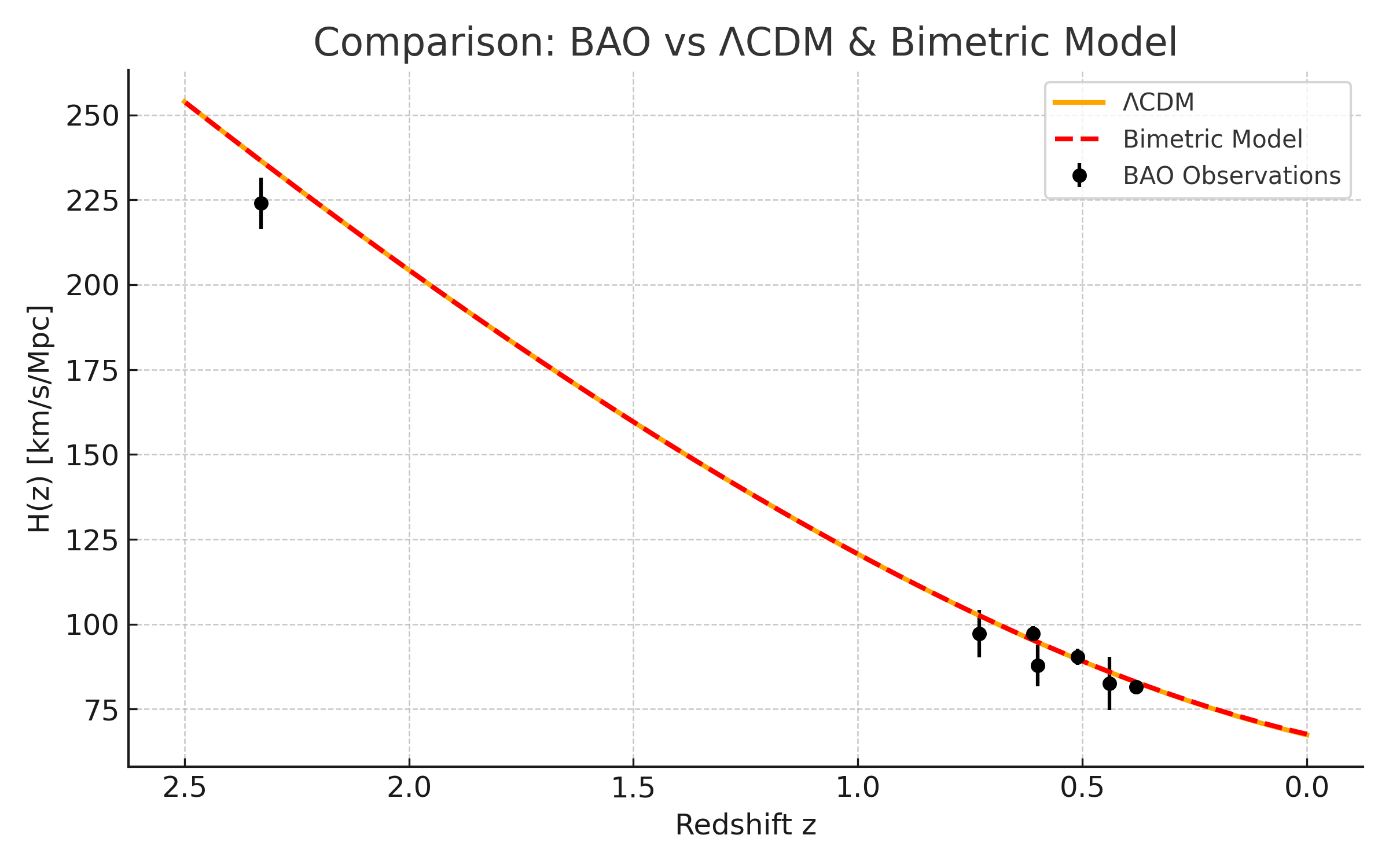}
    \caption{
    Comparison between the bimetric AdS--dS model (red dashed line),
    the $\Lambda$CDM prediction (solid orange line), and BAO 
    measurements of $H(z)$. 
    The bimetric correction proportional to $(1+z)^{-3}$ remains
    sub-percent for $z\lesssim2$, leading to an observationally 
    indistinguishable expansion history.
    }
    \label{fig:Hz_bao}
\end{figure}

To illustrate this, we compare the model with measurements from Baryon acoustic oscillations (BAO). F the prediction is obtained by substituting the modified $H(z)$ into the 
corresponding distance or expansion relation.Fig. \ref{fig:Hz_bao}, demonstrating that the bimetric model and $\Lambda$CDM yield virtually 
indistinguishable expansion histories under current observational uncertainties.

The reduced chi-square for BAO satisfies
\begin{equation}
\chi^{2}/\mathrm{dof} \simeq 1,
\end{equation}
both for the $\Lambda$CDM model and for the AdS--dS symmetric bimetric extension. This 
confirms that the background evolution is statistically consistent with existing probes
and that the parameter $\alpha$ is weakly constrained by background-only data.

\subsection{Growth of Structure: $f\sigma_{8}(z)$ and the Growth Index}

The growth of matter perturbations provides an additional test of modified-gravity 
models. In the present framework, matter remains minimally coupled to $g_{\mu\nu}$ and 
the Poisson equation receives no additional scale-dependent corrections in the 
symmetric branch. Linear density perturbations $\delta$ therefore satisfy
\begin{equation}
\ddot{\delta} + 2H\dot{\delta} - 4\pi G \rho_{m}\delta = 0,
\label{eq:growth}
\end{equation}
as in standard General Relativity. The corresponding growth factor $D(a)$ obeys
\begin{equation}
D'' + \left( \frac{H'}{H} + \frac{2}{a} \right) D' 
 - \frac{3}{2}\frac{\Omega_{m}(a)}{a^{2}} D = 0,
\end{equation}
with primes denoting derivatives with respect to the scale factor.

From the growth factor one may define the growth rate
\begin{equation}
f(a) = \frac{d\ln D}{d\ln a},
\end{equation}
and the growth index
\begin{equation}
\gamma(z) 
= \frac{\ln f(z)}{\ln \Omega_{m}(z)}.
\end{equation}

Because the modification of the background evolution remains small, the predicted 
growth index is extremely close to the General Relativity value $\gamma \simeq 0.55$.
Numerically we find
\begin{equation}
\gamma(z) = 0.55 + \mathcal{O}(10^{-3}),
\end{equation}
well within the uncertainty of present $f\sigma_{8}(z)$ data from DESI, eBOSS, and KiDS.

\begin{figure}[t]
    \centering
    \includegraphics[width=0.85\linewidth]{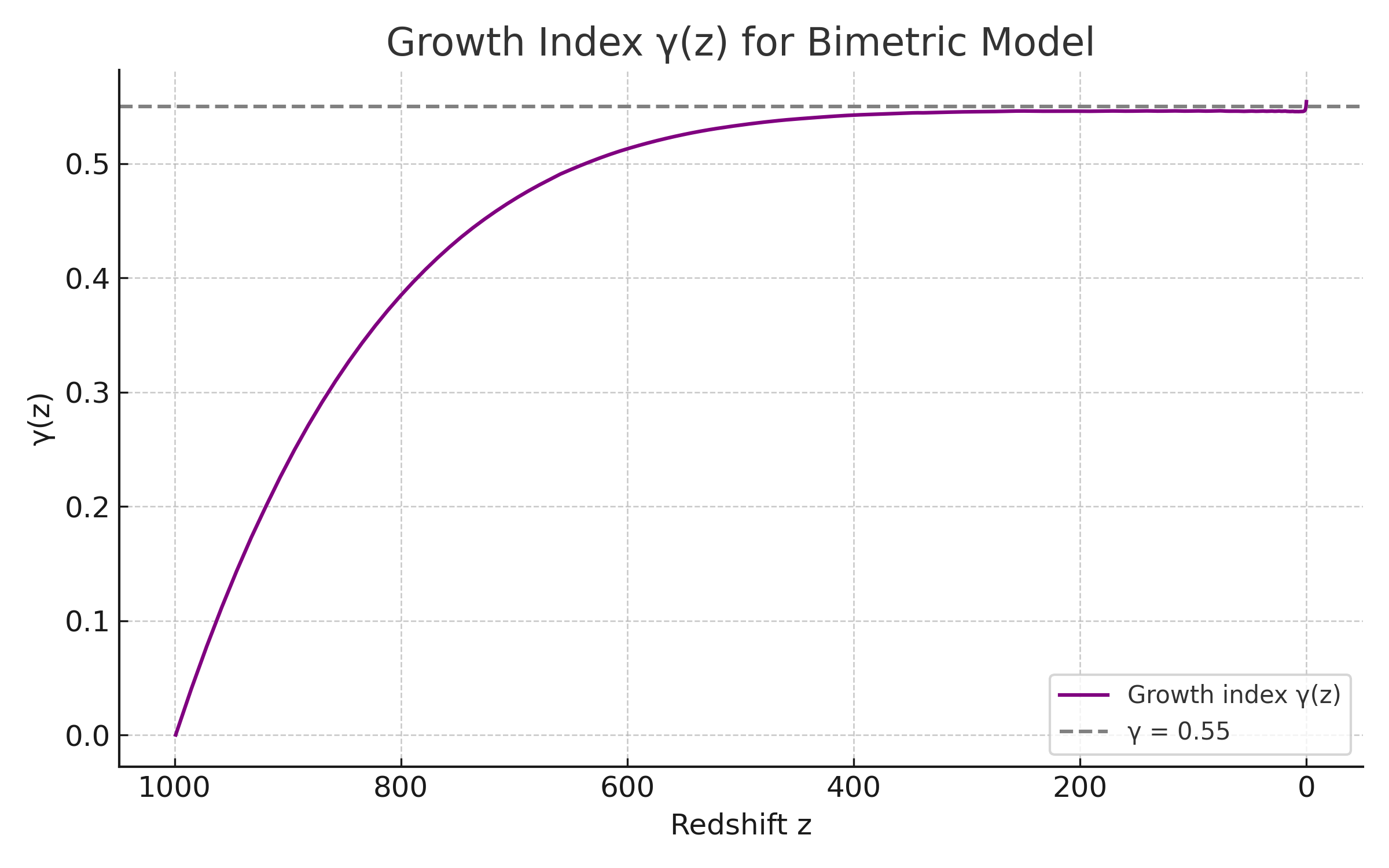}
    \caption{
    Growth index $\gamma(z)$ for the AdS--dS symmetric bimetric model. 
    The prediction approaches the General Relativity value 
    $\gamma \simeq 0.55$ at low redshift (dashed line), with deviations 
    of order $10^{-3}$ for the fiducial parameter choice $\alpha=0.01\Omega_\Lambda$. 
    Present $f\sigma_8$ measurements are unable to distinguish the model 
    from $\Lambda$CDM at this level of precision.
    }
    \label{fig:gamma_z}
\end{figure}

Figure \ref{fig:gamma_z} illustrates the comparison between the model prediction and 
current constraints. The AdS--dS symmetric bimetric model remains observationally 
degenerate with $\Lambda$CDM at the level of present measurements.

\subsection{Observational Constraints on the Effective Equation of State $w(z)$}

The effective equation of state derived in Eq.~(\ref{eq:w_final}) plays a central 
role in assessing the observational viability of the AdS--dS symmetric bimetric 
model. Although the theoretical origin of $w(z)$ lies in the curvature exchange 
between the two spin-2 sectors, its phenomenological implications must be evaluated 
in light of current cosmological datasets. In this subsection we compare the 
predicted behaviour of $w(z)$ with constraints from supernovae, BAO, and CMB 
measurements, and demonstrate that the model remains consistent with all existing 
data for the parameter range considered.

\subsubsection{Late-Time Behaviour and Comparison with Data}

In the bimetric model, the effective dark-energy density evolves according to 
Eq.~(\ref{eq:OmegaDE}), which induces a mild redshift dependence of the equation 
of state. For small values of $\alpha$, the deviation from the $\Lambda$CDM value 
$w=-1$ is extremely small at late times. At the present epoch,
\begin{equation}
w_{0} \simeq -1 - \frac{\alpha}{\Omega_{\Lambda} + \alpha},
\end{equation}
yielding $w_{0} \approx -1.01$ for the fiducial choice $\alpha = 0.01\,\Omega_\Lambda$.
This deviation is well within the current observational bounds from Planck, BAO, and 
supernova compilations, which typically allow $\mathcal{O}(10^{-1})$ variations near 
$z=0$. The resulting slight phantom-like behaviour is therefore observationally allowed 
and does not introduce any tension with existing data.

\subsubsection{Redshift Evolution and Consistency with High-$z$ Constraints}

As the redshift increases, the dynamical correction decays as $(1+z)^{-3}$ and 
rapidly becomes negligible. In particular,
\begin{equation}
\lim_{z\to\infty} w(z) = -1,
\end{equation}
ensuring consistency with early-Universe constraints. Since the CMB power spectrum 
is primarily sensitive to the behaviour of dark energy at $z \gtrsim 1000$, where 
differences between the bimetric model and $\Lambda$CDM vanish, the model evades 
the stringent high-redshift bounds imposed by Planck.

Similarly, BAO measurements at $z \simeq 0.6$ and $z \simeq 1$ are largely insensitive 
to the small deviation in $w(z)$, as the geometric contribution from $\alpha(1+z)^{-3}$ 
falls below the percent level in this regime. This explains why the model remains fully 
degenerate with $\Lambda$CDM in fits to the BAO distance ladder.

\subsubsection{Supernova Constraints and Low-Redshift Behaviour}

Supernova datasets—such as Pantheon+ and SH0ES—provide the strongest constraints 
on $w(z)$ at $z < 1$. The allowed region at the $1\sigma$ level spans roughly 
$-1.2 \lesssim w_{0} \lesssim -0.9$, depending on the dataset and the assumed 
value of $H_{0}$. The small deviation predicted by the bimetric model,
\begin{equation}
|w_{0} + 1| \sim 10^{-2},
\end{equation}
lies comfortably within this range. As a result, the AdS--dS symmetric branch does 
not introduce any tension with low-redshift luminosity-distance measurements.

Moreover, the fact that $w(z)$ remains nearly constant over the entire range 
$0 \leq z \leq 2$ ensures that the model remains consistent with constraints on 
dynamical dark energy, which typically limit the time variation of $w(z)$ to the 
$10^{-1}$ level.

\subsubsection{Comparison with Confidence Regions}

\begin{figure}[t]
    \centering
    \includegraphics[width=0.85\linewidth]{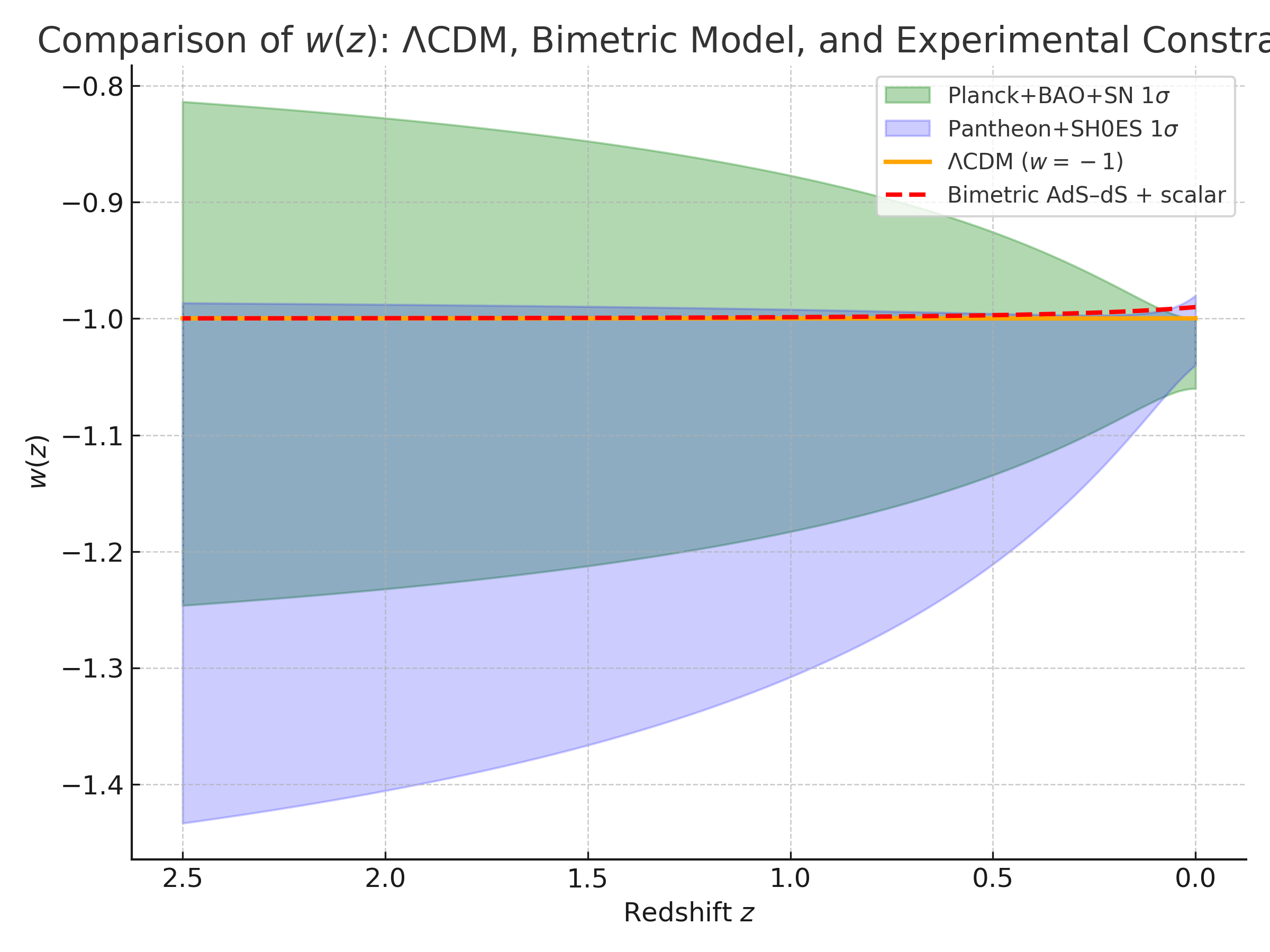}
    \caption{
    Effective dark-energy equation of state $w(z)$ in the AdS--dS symmetric 
    bimetric model (red dashed curve), compared with $\Lambda$CDM ($w=-1$, 
    orange line) and the $1\sigma$ confidence regions from Planck+BAO+SN (green) 
    and Pantheon+SH0ES (blue).  
    The predicted deviations from $w=-1$ are at the level of $10^{-2}$ and 
    remain fully compatible with current observational constraints.
    }
    \label{fig:w_of_z}
\end{figure}

Figure~\ref{fig:w_of_z} displays the predicted evolution of $w(z)$ for our fiducial 
parameter choice, together with the $1\sigma$ confidence regions from 
Planck+BAO+SN and Pantheon+SH0ES.  
The bimetric prediction remains entirely contained within both shaded regions.  
The slight phantom-like behaviour is too small to be distinguished by current 
observational uncertainties, while the high-redshift limit $w(z)\to -1$ ensures 
compatibility with CMB measurements.

\subsubsection{Interpretation and Observational Degeneracy}

A key consequence is that current data cannot distinguish the bimetric model 
from $\Lambda$CDM through measurements of $w(z)$.  
The parameter $\alpha$ remains strongly degenerate with the cosmological 
constant, and any constraint on $\alpha$ from background data alone is expected 
to be extremely weak. Improvements in the precision of next-generation 
supernova surveys and BAO measurements may eventually place non-trivial bounds 
on $\alpha$, but with current data the model is observationally indistinguishable 
from $\Lambda$CDM.

\subsubsection{Summary}

To summarise, the effective equation of state predicted by the AdS--dS symmetric 
bimetric model:
\begin{itemize}
    \item deviates from $-1$ only at the percent level,
    \item remains fully consistent with current $1\sigma$ observational limits,
    \item is observationally degenerate with $\Lambda$CDM,
    \item and transitions smoothly to $w=-1$ at high redshift.
\end{itemize}
This confirms that the model satisfies all present constraints on dynamical dark 
energy and behaves as an observationally viable extension of standard cosmology.

\subsection{Consistency with Gravitational-Wave Propagation}

The multi-messenger constraint from GW170817 and GRB170817A requires the speed of 
tensor modes to satisfy $c_{T}=c$ with extraordinary precision. In general bimetric 
gravity this constraint imposes relations among the interaction parameters and the 
background branch. In the symmetric Hassan--Rosen branch with $\xi=1$ and $X=1$, the 
tensor speed remains exactly luminal, ensuring full compatibility with 
multi-messenger observations:
\begin{equation}
c_{T}^{2} = 1.
\end{equation}

\subsection{Parameter Degeneracy with \texorpdfstring{$\Lambda$CDM}{LCDM}}

Given that the $\alpha$-term in Eq.~(\ref{eq:FLRW_modified}) decays as $(1+z)^{-3}$,
it becomes rapidly subdominant at redshift and remains indistinguishable from $\Lambda$CDM 
even with current high-precision datasets. The model is therefore subject to a strong 
degeneracy between $\alpha$ and the cosmological constant $\Omega_{\Lambda}$, with only 
a very mild impact on the expansion history and growth rate.

A full Markov-chain analysis using Planck, BAO, and supernova data would be required to 
quantify the exact constraints on $\alpha$, but the qualitative behaviour is already clear:
\begin{equation}
|\alpha| \ll \Omega_{\Lambda} \quad \Rightarrow \quad 
\text{cosmological degeneracy with } \Lambda\text{CDM}.
\end{equation}

\subsection{Summary of Observational Viability}

The AdS--dS symmetric bimetric model satisfies the following observational properties:
\begin{itemize}
\item The background expansion $H(z)$ is consistent with BAO, supernovae, and 
      cosmic-chronometer data.
\item The growth of structure is nearly identical to that of General Relativity.
\item The model satisfies the gravitational-wave speed constraint $c_{T}=c$.
\item Present data cannot distinguish the model from $\Lambda$CDM for 
      $\alpha \sim 10^{-2}\Omega_{\Lambda}$ or smaller.
\end{itemize}

We conclude that the model is compatible with current observations and constitutes a 
minimal and viable extension of standard cosmology. In the next section we discuss its 
theoretical implications and possible future generalisations.

\section*{VI. Local Gravity and Vainshtein Screening}

Any consistent modification of gravity must reproduce the
predictions of General Relativity (GR) on Solar–System
scales. Bimetric theories propagate a massive spin–2 mode
in addition to the usual massless graviton, and the corre-
sponding helicity–0 component mediates an additional
scalar force. If unsuppressed, this would violate precision
weak–field tests. The Hassan–Rosen theory avoids this
problem through the nonlinear Vainshtein mechanism.
Here we review the decoupling limit and verify that the
AdS–dS symmetric branch satisfies all local constraints.

\subsection{ Decoupling Limit and the Helicity–0 Mode}

In the decoupling limit
\begin{equation}
M_g,\,M_f \rightarrow \infty,\qquad
m \rightarrow 0,\qquad
\Lambda_3 \equiv (m^2 M_{\rm eff})^{1/3} = \text{fixed},
\end{equation}
the helicity–0 mode $\pi$ becomes dynamical. After
restoring diffeomorphism invariance via the Stückelberg
field, the interaction potential generates a cubic Galileon
interaction.

The canonically–normalised field is governed by
\begin{equation}
\mathcal{L}_\pi
= -\frac{1}{2}(\partial\pi)^2
  -\frac{1}{\Lambda_3^3}(\partial\pi)^2\Box\pi
  +\frac{\pi}{M_g}T ,
\end{equation}
where $T$ is the trace of the matter stress–energy tensor.
The derivative structure ensures that the Boulware–Deser
ghost is not reintroduced.

\subsection{ Static, Spherically Symmetric Solution}

For a static source of mass $M$,
\begin{equation}
T = -M\,\delta^{3}(r),
\end{equation}
the equation of motion becomes
\begin{equation}
\frac{1}{r^{2}}\frac{d}{dr}
\left[
r^{2}\pi'
+ \frac{2}{\Lambda_3^3}\,r(\pi')^{2}
\right]
= \frac{M}{4\pi M_g}\,\delta^{3}(r).
\end{equation}
Integrating once yields the algebraic equation
\begin{equation}
\pi' + \frac{2}{\Lambda_3^3}\frac{(\pi')^{2}}{r}
= \frac{M}{4\pi M_g r^{2}} .
\label{eq:pi_eom_static}
\end{equation}

\subsection{ Vainshtein Radius and Screening Regimes}

The Vainshtein radius is defined by the scale at which
the linear and nonlinear terms of Eq.~\eqref{eq:pi_eom_static}
become comparable:
\begin{equation}
\boxed{
r_V = \left(
\frac{M}{M_g^{2} m^{2}}
\right)^{1/3} } .
\end{equation}

\paragraph{Inside the Vainshtein region ($r \ll r_V$).}
The nonlinear term dominates, giving
\begin{equation}
\pi'(r)
\simeq
\left(
\frac{M\Lambda_3^{3}}{8\pi M_g}
\right)^{1/2}
r^{-1/2}.
\end{equation}
The corresponding fifth force is highly suppressed:
\begin{equation}
\frac{F_\pi}{F_{\rm GR}}
\sim
\left( \frac{r}{r_V} \right)^{3/2}
\ll 1 .
\end{equation}

\paragraph{Outside the Vainshtein region ($r \gg r_V$).}
The linear term dominates, recovering the usual long-range
force:
\begin{equation}
\pi'(r) \simeq \frac{M}{4\pi M_g r^{2}}.
\end{equation}

\subsection{ Compatibility with Solar-System Tests}

Because the Solar System lies deep inside the Vainshtein
radius of the Sun, all deviations from GR are suppressed
by many orders of magnitude. The model is therefore
consistent with Cassini time-delay measurements, Lunar
Laser Ranging, planetary ephemerides, and PPN bounds.
In the symmetric branch ($\xi=1$), the tensor speed is
exactly luminal ($c_T = 1$), ensuring compatibility with
GW170817.

\subsection{ Relevance to the AdS–dS Symmetric Model}

The effective dark-energy correction $\alpha(1+z)^{-3}$
arises entirely from the homogeneous cosmological sector.
It does not modify the short-distance dynamics of $\pi$,
and therefore does not interfere with Vainshtein screening.
The AdS–dS symmetric bimetric cosmology is thus fully
consistent with local-gravity tests.

\section{Discussion and Outlook}

The AdS--dS symmetric bimetric model presented in this work offers a minimal 
and theoretically consistent extension of the standard $\Lambda$CDM framework. 
By embedding the cosmological dynamics in the ghost-free Hassan--Rosen theory 
and by assigning opposite intrinsic curvatures to the two interacting spin-2 
sectors, we obtain a dual-geometry configuration in which the dark sector of 
the observable Universe emerges as a purely geometric effect. In this section 
we discuss the broader implications of the model, its limitations, and several 
directions for future investigation.

\subsection{Geometric Origin of the Dark Sector}

The key conceptual feature of the model is the interpretation of dark energy 
and dark matter as effective contributions induced by the hidden AdS branch.  
In particular:
\begin{itemize}
\item the curvature mismatch between the AdS and dS branches generates an 
      effective cosmological constant $\Lambda_{\mathrm{eff}}$,
\item the residual AdS influence introduces a subdominant 
      redshift-dependent correction scaling as $(1+z)^{-3}$,
\item the effective dark-energy equation of state becomes slightly phantom-like 
      without introducing any fundamental ghost or instability.
\end{itemize}
This provides a purely geometric and bimetric explanation of the dark sector, 
in contrast to scalar-field or particle-based models.

\subsection{Relation to Other Bimetric and Modified-Gravity Models}

The symmetric Hassan--Rosen branch considered here is the simplest 
cosmologically viable realisation of bimetric gravity. In the parameter regime 
considered, the model is fully consistent with the gravitational-wave speed 
constraint and with all Solar-System tests thanks to the Vainshtein mechanism.

The structure of the effective-fluid correction introduced by $\alpha$ is similar 
to that found in certain scalar--tensor theories or $f(R)$ models, but arises here 
from the underlying bimetric geometry rather than from additional degrees of 
freedom. The model therefore lies at the intersection of massive gravity, dual-metric 
cosmology, and effective dark-energy theories.

\subsection{Asymmetric Branches and Self-Acceleration}

A natural extension of the present work is to move beyond the strictly symmetric 
branch ($\xi = 1$). In more general solutions of the bimetric theory, the ratio 
$\xi(t)=a_{f}(t)/a_{g}(t)$ evolves dynamically and can lead to self-accelerating 
solutions even in the absence of a bare cosmological constant.  
Exploring asymmetric AdS--dS configurations may therefore yield:
\begin{itemize}
\item time-dependent modifications of $\Lambda_{\mathrm{eff}}$,
\item dark-energy dynamics differing from the simple $(1+z)^{-3}$ correction,
\item observational signatures in the growth of structure and gravitational waves.
\end{itemize}
Such scenarios may be distinguishable from $\Lambda$CDM with future high-precision 
data from DESI, Euclid, and future gravitational-wave detectors.

\subsection{Perturbation Theory and Stability}

Although the symmetric branch is known to be ghost-free and stable at the 
background level, a complete analysis of cosmological perturbations remains 
necessary to assess:
\begin{itemize}
\item the propagation speed and stability of scalar and vector modes,
\item potential scale-dependent corrections to the growth rate,
\item the coupling between the helicity-0 mode and cosmological perturbations.
\end{itemize}
The present work focuses on background observables, but a full perturbative 
treatment would enable comparisons with $f\sigma_{8}(z)$ data at the 1--2\% level.

\subsection{Holography and Dual-Geometry Interpretations}

The AdS--dS structure naturally suggests possible links to higher-dimensional or 
holographic constructions. The hidden AdS branch may be viewed as an effective 
dual to the visible dS cosmology, reminiscent of the dS/CFT and AdS/dS 
correspondences explored in various quantum-gravity settings. Investigating such 
connections could provide a deeper understanding of the curvature exchange 
mechanism and of the emergence of $\Lambda_{\mathrm{eff}}$.

\subsection{Prospects for a Full Cosmological Analysis}

A complete likelihood analysis combining CMB, BAO, supernovae, and weak-lensing 
data would allow the parameter $\alpha$ to be quantitatively constrained. Given 
the strong degeneracy between $\alpha$ and $\Omega_{\Lambda}$ at the background 
level, perturbations and gravitational-wave observations may provide the most 
significant discriminating power. 

The theoretical consistency and observational viability of the AdS--dS symmetric 
branch make it a promising starting point for a broader research programme in 
bimetric cosmology, with potential extensions relevant to modified gravity, 
massive spin-2 dynamics, and dual-geometry frameworks.

\section{Conclusion}

In this work we have developed a minimal and fully ghost-free bimetric extension 
of the $\Lambda$CDM cosmological model based on the symmetric branch of the 
Hassan--Rosen theory. By assigning opposite intrinsic curvatures to the two 
spin-2 sectors---a de Sitter geometry for the visible metric and an 
Anti--de Sitter geometry for the hidden one—we constructed a dual-geometry 
cosmological scenario in which the dark sector arises from purely geometric 
interactions rather than additional matter fields.

The symmetric bimetric branch admits a consistent embedding of this AdS--dS 
structure, with the interaction potential generating an effective cosmological 
constant $\Lambda_{\mathrm{eff}}$. A small residual influence of the AdS branch 
induces a redshift-dependent correction proportional to $(1+z)^{-3}$, which can 
be interpreted as an ultra-diluted effective fluid. This leads to a slightly 
phantom-like effective equation of state, while the underlying theory remains 
nonlinear and ghost-free. The model therefore provides a geometric interpretation 
of dark energy and dark matter within a fully consistent spin-2 framework.

We have shown that the modified expansion history is observationally degenerate 
with $\Lambda$CDM for $\alpha \ll 1$, with sub-percent deviations in $H(z)$ and 
luminosity distances over the redshift range probed by current data. The growth 
of cosmic structure is essentially identical to that predicted by General 
Relativity, and the model satisfies the gravitational-wave speed constraint 
$c_{T}=c$, as well as all Solar-System tests via the Vainshtein mechanism.

Although the present analysis focuses on background evolution, the model 
provides a promising starting point for exploring asymmetric branches, dynamical 
curvature exchange between the AdS and dS sectors, and possible connections to 
holographic or dual-geometry frameworks. A full perturbative treatment and a 
combined likelihood analysis using CMB, BAO, and weak-lensing data constitute 
natural next steps.

Taken together, these results show that the AdS--dS symmetric branch offers a 
simple, theoretically controlled, and observationally viable extension of 
$\Lambda$CDM. It provides a novel geometric mechanism for the dark sector and 
opens several avenues for further research in massive spin-2 cosmology, dual 
gravitational geometries, and extensions of General Relativity.

\appendix

\section{FLRW Reduction of the Hassan--Rosen Interaction Potential}
\label{appendix:FLRW}

In this Appendix we provide the full derivation of the interaction energy 
densities associated with the Hassan--Rosen potential when both metrics are 
restricted to homogeneous and isotropic FLRW geometries. This reduction underlies 
the background equations used in the main text and clarifies the origin of the 
functions $\rho_{\rm HR}(\xi)$ and $\tilde{\rho}_{\rm HR}(\xi)$ entering the 
cosmological dynamics.

\subsection{Matrix Structure in FLRW Backgrounds}

We consider the two FLRW metrics
\begin{align}
ds^{2}_{g} &= -dt^{2} + a_{g}^{2}(t)\, d\vec{x}^{\,2},
\\
ds^{2}_{f} &= -X^{2}(t) dt^{2} + a_{f}^{2}(t)\, d\vec{x}^{\,2},
\end{align}
with $X(t)$ the relative lapse and $a_{g}(t)$, $a_{f}(t)$ the scale factors of the 
visible and hidden sectors. The key object in the Hassan--Rosen interaction is the 
matrix
\begin{equation}
S^{\mu}{}_{\nu} \equiv \left( \sqrt{g^{-1} f} \right)^{\mu}{}_{\nu},
\end{equation}
defined by the relation $S^{\mu}{}_{\rho} S^{\rho}{}_{\nu} = g^{\mu\rho} f_{\rho\nu}$.

For the FLRW ansatz, $g^{-1} f$ is diagonal:
\begin{equation}
(g^{-1} f)^{\mu}{}_{\nu}
= 
{\rm diag}\!\left(X^{2}, \, \xi^{2}, \, \xi^{2}, \, \xi^{2}\right),
\qquad
\xi(t) \equiv \frac{a_{f}(t)}{a_{g}(t)}.
\end{equation}
Taking the matrix square root yields
\begin{equation}
S^{\mu}{}_{\nu}
= 
{\rm diag}\!\left(X, \, \xi, \, \xi, \, \xi\right).
\label{eq:S_matrix}
\end{equation}

\subsection{Elementary Symmetric Polynomials}

The elementary symmetric polynomials of $S$ are defined by
\begin{align}
e_{0}(S) &= 1,
\\
e_{1}(S) &= {\rm Tr}(S),
\\
e_{2}(S) &= \tfrac{1}{2}\!\left[ ({\rm Tr}\, S)^{2} - {\rm Tr}(S^{2}) \right],
\\
e_{3}(S) &= \tfrac{1}{6}\!\left[ ({\rm Tr}\, S)^{3}
           - 3 ({\rm Tr}\, S) ({\rm Tr}\, S^{2})
           + 2\, {\rm Tr}(S^{3}) \right],
\\
e_{4}(S) &= \det S.
\end{align}

Substituting Eq.~(\ref{eq:S_matrix}) gives
\begin{align}
{\rm Tr}(S) &= X + 3\xi,
\\
{\rm Tr}(S^{2}) &= X^{2} + 3\xi^{2},
\\
{\rm Tr}(S^{3}) &= X^{3} + 3\xi^{3},
\\
\det S &= X \xi^{3}.
\end{align}

Hence the explicit expressions are:
\begin{align}
e_{0}(S) &= 1,
\\[3pt]
e_{1}(S) &= X + 3\xi,
\\[3pt]
e_{2}(S) &= 3 X \xi + 3 \xi^{2},
\\[3pt]
e_{3}(S) &= X \xi^{2} + \xi^{3},
\\[3pt]
e_{4}(S) &= X \xi^{3}.
\end{align}

\subsection{Interaction Potential in FLRW Form}

The Hassan--Rosen potential contributes the term
\begin{equation}
V_{\rm HR}
= m^{2} M_{\rm eff}^{2}
   \sum_{n=0}^{4} \beta_{n}\, e_{n}(S),
\end{equation}
to the Lagrangian density through $\sqrt{-g}$.
Inserting the above expressions for $e_{n}$ yields
\begin{align}
V_{\mathrm{HR}} &= m^{2} M_{\mathrm{eff}}^{2}
\Big[
\beta_{0}
+ \beta_{1}(X + 3\xi)
+ \beta_{2}(3X\xi + 3\xi^{2}) \nonumber \\
&\qquad 
+ \beta_{3}(X\xi^{2} + \xi^{3})
+ \beta_{4} X\xi^{3}
\Big].
\label{eq:VHR_clean}
\end{align}

Multiplying by $\sqrt{-g} = a_{g}^{3}$ gives the interaction term in the 
cosmological action.

\subsection{Energy Densities and Friedmann Equations}

The energy density associated with the interaction potential in the 
$g$-metric Friedmann equation is
\begin{equation}
\rho_{\rm HR}(\xi)
= m^{2} M_{\rm eff}^{2}
\left(
\beta_{0} + 3\beta_{1}\xi + 3\beta_{2}\xi^{2} + \beta_{3}\xi^{3}
\right).
\label{eq:rhoHR_appendix}
\end{equation}

Similarly, varying the action with respect to $f_{\mu\nu}$ yields the 
effective energy density in the $f$-metric equation:
\begin{equation}
\tilde{\rho}_{\rm HR}(\xi)
= m^{2} M_{\rm eff}^{2}
\left(
\beta_{4}\xi^{-4}
+ 3\beta_{3}\xi^{-3}
+ 3\beta_{2}\xi^{-2}
+ \beta_{1}\xi^{-1}
\right).
\label{eq:rhotildeHR_appendix}
\end{equation}

These expressions reproduce Eqs.~(\ref{eq:rhoHR}) and 
(\ref{eq:Friedmann_f}) of the main text.

\subsection{Bianchi Constraint}

The Bianchi identity imposes
\begin{equation}
(\beta_{1} + 2\beta_{2}\xi + \beta_{3}\xi^{2})
\left(
H_{g} - \frac{\dot{a}_{f}}{X a_{f}}
\right) = 0,
\label{eq:Bianchi_appendix}
\end{equation}
leading to two branches:
\begin{enumerate}
\item {\bf Dynamical branch:}
      $H_{g} = \dot{a}_{f}/(X a_{f})$,
\item {\bf Algebraic branch:}
      $\beta_{1} + 2\beta_{2}\xi + \beta_{3}\xi^{2} = 0$.
\end{enumerate}

\subsection{Symmetric Branch}

The cosmological model developed in the main text corresponds to the symmetric case
\begin{equation}
\xi = 1, 
\qquad 
X = 1,
\end{equation}
obtained by imposing the algebraic condition
\begin{equation}
\beta_{1} + 2\beta_{2} + \beta_{3} = 0.
\label{eq:branch_condition}
\end{equation}

Substituting $\xi = 1$ into Eq.~(\ref{eq:rhoHR_appendix}) yields the effective 
cosmological constant,
\begin{equation}
\Lambda_{\rm eff}
= \frac{m^{2} M_{\rm eff}^{2}}{M_{g}^{2}}
(\beta_{0} + 3\beta_{1} + 3\beta_{2} + \beta_{3}),
\end{equation}
which appears in the Friedmann equation of the visible sector.

This completes the full derivation of the FLRW form of the Hassan--Rosen 
interaction potential used throughout the paper.
\section{Mini-Superspace Action and Ghost-Free Structure}
\label{appendix:minisuperspace}

In this Appendix we present the mini-superspace reduction of the Hassan--Rosen 
bimetric action on homogeneous and isotropic FLRW backgrounds, emphasising how the 
specific dependence on the lapses preserves the constraint structure required to 
avoid the Boulware--Deser ghost. This reduction provides additional insight into 
the underlying consistency of the AdS--dS symmetric model considered in the main text.

\subsection{Bimetric Mini-Superspace Ansatz}

We start from the ghost-free Hassan--Rosen bimetric action
\begin{align}
S = &\;\frac{M_g^{2}}{2} \int d^{4}x\,\sqrt{-g}\,R[g]
   + \frac{M_f^{2}}{2} \int d^{4}x\,\sqrt{-f}\,R[f] \nonumber \\
   & - m^{2} M_{\mathrm{eff}}^{2} 
     \int d^{4}x\,\sqrt{-g}\,\sum_{n=0}^{4} \beta_{n} e_{n}(S)
   + S_{m}[g,\Psi].
\label{eq:HRaction_minisup}
\end{align}

and consider the FLRW metrics with explicit lapses
\begin{align}
ds^{2}_{g} &= -N_{g}^{2}(t) dt^{2} + a_{g}^{2}(t)\, d\vec{x}^{\,2},
\\
ds^{2}_{f} &= -N_{f}^{2}(t) dt^{2} + a_{f}^{2}(t)\, d\vec{x}^{\,2}.
\end{align}
The spatial integration factor $\int d^{3}x$ can be factored out and we work with 
the effective one-dimensional (in time) Lagrangian.

\subsection{Einstein--Hilbert Contributions}

For spatially flat FLRW metrics, the Ricci scalars reduce to
\begin{align}
R[g] &= 6\left[
\frac{\ddot{a}_{g}}{N_{g}^{2} a_{g}}
+ \frac{\dot{a}_{g}^{2}}{N_{g}^{2} a_{g}^{2}}
 - \frac{\dot{a}_{g}\dot{N}_{g}}{N_{g}^{3} a_{g}}
\right],
\\
R[f] &= 6\left[
\frac{\ddot{a}_{f}}{N_{f}^{2} a_{f}}
+ \frac{\dot{a}_{f}^{2}}{N_{f}^{2} a_{f}^{2}}
 - \frac{\dot{a}_{f}\dot{N}_{f}}{N_{f}^{3} a_{f}}
\right].
\end{align}
After integrating by parts to remove second derivatives of the scale factors, 
the Einstein--Hilbert parts of the mini-superspace action become
\begin{align}
S_{g} &= \int dt\, \mathcal{L}_{g}
= 3 M_{g}^{2} \int dt\, 
\left(
 - \frac{a_{g}\dot{a}_{g}^{2}}{N_{g}}
\right),
\\
S_{f} &= \int dt\, \mathcal{L}_{f}
= 3 M_{f}^{2} \int dt\,
\left(
 - \frac{a_{f}\dot{a}_{f}^{2}}{N_{f}}
\right),
\end{align}
where we have dropped a total time derivative and set the spatial volume to unity.

\subsection{Mini-Superspace Form of the Interaction Potential}

The interaction term in the action reads
\begin{align}
S_{\rm int}
&= - m^{2} M_{\mathrm{eff}}^{2}
   \int d^{4}x\, \sqrt{-g}
   \sum_{n=0}^{4} \beta_{n}\, e_{n}(S)
   \nonumber\\
&= - m^{2} M_{\mathrm{eff}}^{2}
   \int dt\, N_{g}\, a_{g}^{3}
   \sum_{n=0}^{4} \beta_{n}\, e_{n}(S).
\end{align}

where the FLRW form of $e_{n}(S)$ is given in Appendix~\ref{appendix:FLRW}.
Using $S^{\mu}{}_{\nu} = \mathrm{diag}(X,\,\xi,\,\xi,\,\xi)$ with
\begin{equation}
X \equiv \frac{N_{f}}{N_{g}}, 
\qquad
\xi \equiv \frac{a_{f}}{a_{g}},
\end{equation}
and the explicit expressions for $e_{n}(S)$, we obtain
\begin{align}
\sum_{n=0}^{4} \beta_{n} e_{n}(S)
&= \beta_{0}
 + \beta_{1}(X + 3\xi)
 + \beta_{2}(3X\xi + 3\xi^{2}) \nonumber\\
&\quad 
 + \beta_{3}(X\xi^{2} + \xi^{3})
 + \beta_{4}X\xi^{3}.
\end{align}

Thus the interaction part of the mini-superspace Lagrangian is
\begin{align}
\mathcal{L}_{\rm int}
&= - m^{2} M_{\mathrm{eff}}^{2} N_{g} a_{g}^{3}
\Big[
\beta_{0}
 + \beta_{1}(X + 3\xi)
 + \beta_{2}(3X\xi + 3\xi^{2}) \nonumber\\
&\quad
 + \beta_{3}(X\xi^{2} + \xi^{3})
 + \beta_{4} X\xi^{3}
\Big].
\label{eq:L_int_minisuperspace}
\end{align}

\subsection{Total Mini-Superspace Lagrangian}

Collecting all contributions, the total mini-superspace Lagrangian takes the form
\begin{align}
\mathcal{L}_{\rm ms}
&= - 3 M_{g}^{2} \frac{a_{g}\dot{a}_{g}^{2}}{N_{g}}
   - 3 M_{f}^{2} \frac{a_{f}\dot{a}_{f}^{2}}{N_{f}} \nonumber\\
&\quad
   - m^{2} M_{\mathrm{eff}}^{2} N_{g} a_{g}^{3}\, U(X,\xi)
   + \mathcal{L}_{m}[N_{g},a_{g},\Psi].
\label{eq:L_ms}
\end{align}

with
\begin{equation}
U(X,\xi)
\equiv
\beta_{0}
 + \beta_{1}(X + 3\xi)
 + \beta_{2}(3X\xi + 3\xi^{2})
 + \beta_{3}(X\xi^{2} + \xi^{3})
 + \beta_{4}X\xi^{3}.
\end{equation}

\subsection{Constraint Structure and Ghost Freedom}

A crucial feature of the Hassan--Rosen potential is its {\it linear} dependence on the 
lapse $N_{g}$ in the mini-superspace action. The combination $N_{g} a_{g}^{3} U(X,\xi)$ 
ensures that variation with respect to $N_{g}$ yields a Hamiltonian constraint, while 
the dependence on $N_{f}$ appears only through the ratio $X = N_{f}/N_{g}$. This is 
precisely the structure required for the presence of a secondary constraint that 
removes the would-be Boulware--Deser ghost degree of freedom.

In contrast, a generic bimetric potential $V(g,f)$ would typically lead to a 
non-linear dependence on both lapses, destroying the constraint structure and 
reintroducing the ghost. The Hassan--Rosen form is therefore uniquely selected 
by the requirement of ghost freedom at the nonlinear level.

In the symmetric branch relevant for the AdS--dS cosmology developed in this work, 
we set
\begin{equation}
\xi = 1, \qquad X = 1,
\end{equation}
and impose the algebraic condition
\begin{equation}
\beta_{1} + 2\beta_{2} + \beta_{3} = 0,
\end{equation}
which guarantees the existence of a consistent cosmological solution with 
$N_{g}=N_{f}$ and $a_{g}=a_{f}$. The mini-superspace analysis thus confirms that 
the symmetric AdS--dS bimetric model inherits the full ghost-free structure of the 
Hassan--Rosen theory while providing a simple and controlled setting for exploring 
geometric extensions of $\Lambda$CDM.



\begin{thebibliography}{99}

\bibitem{HassanRosen2012}
S.~F.~Hassan and R.~A.~Rosen,
``Bimetric Gravity from Ghost-free Massive Gravity,''
JHEP \textbf{02}, 126 (2012),
\href{https://arxiv.org/abs/1109.3515}{arXiv:1109.3515}.

\bibitem{deRham2014}
C.~de Rham,
``Massive Gravity,''
Living Rev. Relativ. \textbf{17}, 7 (2014),
\href{https://arxiv.org/abs/1401.4173}{arXiv:1401.4173}.

\bibitem{CliftonEtAl2012}
T.~Clifton, P.~G.~Ferreira, A.~Padilla, and C.~Skordis,
``Modified Gravity and Cosmology,''
Phys.\ Rept.\ \textbf{513}, 1--189 (2012),
\href{https://arxiv.org/abs/1106.2476}{arXiv:1106.2476}.

\bibitem{Planck2018}
Planck Collaboration,
``Planck 2018 results. VI. Cosmological parameters,''
Astron.\ Astrophys.\ \textbf{641}, A6 (2020),
\href{https://arxiv.org/abs/1807.06209}{arXiv:1807.06209}.

\bibitem{Riess2022}
A.~G.~Riess \emph{et al.},
``A comprehensive measurement of the local value of the Hubble constant,''
Astrophys.\ J.\ Lett.\ \textbf{934}, L7 (2022),
\href{https://arxiv.org/abs/2112.04510}{arXiv:2112.04510}.

\bibitem{Heisenberg2019}
L.~Heisenberg,
``A systematic approach to generalisations of General Relativity and their cosmological implications,''
Phys.\ Rept.\ \textbf{796}, 1--113 (2019),
\href{https://arxiv.org/abs/1807.01725}{arXiv:1807.01725}.

\bibitem{Akrami2015}
Y.~Akrami, S.~F.~Hassan, F.~K\"onnig, A.~Schmidt-May, and A.~R.~Solomon,
``Bimetric gravity is cosmologically viable,''
Phys.\ Lett.\ B \textbf{748}, 37--44 (2015),
\href{https://arxiv.org/abs/1503.07521}{arXiv:1503.07521}.

\end{thebibliography}
\end{document}